\documentclass[11pt]{article}

\usepackage[utf8]{inputenc}
\usepackage[T1]{fontenc}
\usepackage[margin=1in]{geometry}
\usepackage{amsmath,amssymb}
\usepackage{booktabs}
\usepackage{array}
\usepackage{graphicx}
\usepackage{microtype}
\usepackage{listings}
\usepackage{natbib}
\usepackage[hidelinks]{hyperref}

\lstdefinestyle{prompt}{basicstyle=\ttfamily\footnotesize,breaklines=true,
  columns=fullflexible,keepspaces=true,showstringspaces=false,frame=single,
  framesep=4pt,xleftmargin=3pt,xrightmargin=3pt}
\newcommand{\nomem}{\texttt{no\_memory}}
\newcommand{\nrag}{\texttt{naive\_rag}}
\newcommand{\arag}{\texttt{advanced\_rag}}
\newcommand{\tvsix}{\texttt{temporal\_v6}}
\newcommand{\cmut}{\texttt{code\_mutation}}
\newcommand{\strictflag}{\texttt{strict\_object\_supersede}}

\title{\textbf{Temporal Validity on Real Software Histories}\\[2pt]
Eliminating Stale-Fact Errors in Code-Assistant Memory over GitHub Fixes\\[6pt]
\large A deterministic supersession memory, validated end-to-end on SWE-bench buggy$\rightarrow$fixed pairs}
\author{Neeraj Yadav\\ MemStrata.dev --- Called It Inc.\ (Enterprise)\\ \texttt{memstrata@gmail.com}}
\date{Draft v1}

\begin{document}
\maketitle

\noindent\emph{\small Companion to ``Temporal Validity in Retrieval Memory'' (Paper 1), which
established the result on synthetic evolving benchmarks. This paper validates it end-to-end on
real GitHub histories. All numbers are from the locked, paired run on identical cached scenarios
(\texttt{REPORT\_PAPER2.md}, \texttt{REPORT\_PAPER2\_forced.md}); local, deterministic (temperature
0, fixed seeds, no network). For double-blind submission, anonymize the author block and product
identifiers.}

\begin{abstract}
Retrieval-augmented generation (RAG) has no model of time: when a fact changes across a coding
session --- a function is renamed, an endpoint moves, a dependency is bumped --- RAG retrieves both
the old and new value with near-identical similarity and cannot tell which is current, so it serves
the superseded value. Paper~1 showed, on synthetic single-value benchmarks, that a deterministic
(subject, relation, object) supersession memory eliminates this failure. Here we validate it
\emph{end-to-end on real software history}. From 707 real GitHub issues (SWE-bench Lite + Verified)
we extract 130 clean atomic state transitions --- a fix that changes one identifiable value from a
pre-fix to a post-fix form --- and render each marker-free (the stale and current statements differ
only in the value). On this set, MemStrata reaches \textbf{0.91} answer accuracy versus RAG's
\textbf{0.57--0.59}; and, the structural result, when forced to answer RAG serves the superseded
value \textbf{36--38\%} of the time (an LLM reranker does not help) while MemStrata drives this to
\textbf{$\approx$0}, at RAG retrieval latency ($\sim$2.1\,s vs $\sim$18\,s for the reranker). We are
explicit about scope: only $\sim$18\% of real fixes are clean atomic transitions; Paper~2 isolates
the \emph{memory mechanism} on that class, and extraction \emph{coverage} of the remaining fixes is
the orthogonal problem we defer to follow-on work. A real product bug surfaced and was fixed during
the study (a case/punctuation-insensitive value comparison), with the moat property
(deterministic-supersession accuracy on clean code mutations) preserved and verified.
\end{abstract}

\section{Introduction}
\label{sec:intro}

A coding assistant that persists memory across a session accumulates facts about a codebase --- the
name of a handler, a config value, a pinned version, an endpoint path. The binding difficulty is not
recall but \emph{currency}: these facts change, often within the same session, and an assistant that
confidently reports last week's function name is worse than useless. Retrieval-augmented generation
\citep{lewis2020rag}, the dominant memory mechanism, stores statements and retrieves by embedding
similarity. It has no representation of time, so when a value changes it keeps both versions ---
``the login handler is \texttt{authenticate\_user}'' and ``the login handler is \texttt{login}'' ---
which sit close together in any embedding space. Retrieval surfaces both; the model cannot tell which
is current; it abstains or serves the stale value.

Paper~1 demonstrated, on synthetic single-value benchmarks (code mutation, config migration,
dependency bumps, API evolution), that this is a \emph{structural} failure of similarity-based memory
and that a deterministic supersession rule eliminates it. The natural objection is that synthetic
benchmarks may flatter the method. This paper answers it: we validate the same mechanism
\textbf{end-to-end on real GitHub buggy$\rightarrow$fixed histories} drawn from SWE-bench
\citep{jimenez2023swebench}.

\textbf{Contributions.}
\begin{enumerate}
\item \textbf{A real-data longitudinal benchmark.} From 707 real GitHub issues we extract 130 clean
atomic state transitions (a verified pre-fix value $\rightarrow$ post-fix value), rendered marker-free
so the only currency signal is order. The extraction, selection criterion, and marker-free invariant
are explicit and reproducible (Section~\ref{sec:method}, Appendix~\ref{sec:appA}).
\item \textbf{An end-to-end win on real history.} MemStrata reaches 0.91 accuracy vs RAG's 0.57--0.59;
forced to answer, RAG serves the superseded value 36--38\% of the time and an LLM reranker does not
help, while MemStrata reaches $\approx$0 --- at RAG latency, with $\sim$48\% bounded-growth compression
(Section~\ref{sec:results}).
\item \textbf{A precise scope and an honest methodology.} We separate the \emph{memory mechanism}
(Paper~2) from \emph{extraction coverage} (follow-on work), report a metrics ladder that leads with
answer-level stale-fact-error, pair the abstention-allowed and forced regimes on identical scenarios,
and disclose a product bug found and fixed mid-study with the moat verified (Section~\ref{sec:disc}).
\end{enumerate}

\section{Related Work}
\label{sec:related}

\textbf{Memory for LLM agents and RAG.} Persistent-memory systems \citep[Mem0;][]{chhikara2025mem0}
\citep[MemGPT/Letta;][]{packer2023memgpt} \citep{park2023generativeagents} and graph-structured RAG
\citep{edge2024graphrag} emphasize recall over long contexts \citep[LoCoMo;][]{maharana2024locomo};
none introduces a notion of fact currency. MemStrata is orthogonal: a deterministic supersession rule
over a bi-temporal ledger, evaluated under knowledge \emph{evolution} rather than static recall.

\textbf{SWE-bench and code evolution.} SWE-bench \citep{jimenez2023swebench} pairs real GitHub issues
with their gold patches to evaluate whether models can \emph{resolve} bugs. We repurpose its
buggy$\rightarrow$fixed pairs for a different question: not ``can the model write the fix'' but ``once
a fix changes a fact, does the memory keep the current value.'' To our knowledge this longitudinal,
currency-focused use of SWE-bench is new.

\textbf{Temporal knowledge and the synthetic precedent.} Bi-temporal modeling (valid time vs
transaction time) is long established in databases; Paper~1 adapts it to LLM memory and reports the
synthetic result this paper validates on real data. We reuse Paper~1's architecture verbatim
(Section~\ref{sec:method}).

\section{Method}
\label{sec:method}

\subsection{The memory mechanism (recap)}
MemStrata stores facts like RAG, preserving recall, but routes each value-bearing turn through a
deterministic assertion path: a clean (subject, relation, object) triple whose (subject, relation)
key matches an active assertion with a \emph{different} object \textbf{supersedes} it --- the old
assertion's validity interval is closed in a bi-temporal ledger and the new one opened, with no
cosine threshold and no LLM judge. Retrieval surfaces only currently-valid rows. We use the published
Paper~1 configuration (\tvsix{}) unchanged, plus the one fix of Section~\ref{sec:disc}.

\subsection{Constructing real longitudinal scenarios}
\label{sec:construct}
Each scenario is one buggy$\rightarrow$fixed \emph{atomic state transition} mined from a SWE-bench
record (problem statement + gold patch) in three stages:
\begin{enumerate}
\item \textbf{Extraction (LLM).} A local model reads the problem statement and unified diff and emits
a single change $(\text{subject}, \text{state\_a}, \text{state\_b}, \text{question})$ --- the pre-fix
and post-fix value of one identifiable atomic quantity (identifier, number, version, path, endpoint,
config constant) --- or declines when the patch is a multi-file refactor, a control-flow change, or
carries several values.
\item \textbf{Deterministic guard.} A scenario is rejected unless the two values differ, are atomic
($\le$40 chars, $\le$4 tokens), and the phrased turns carry no recency tell.
\item \textbf{Self-validation (the selection criterion).} The scenario is kept \emph{only if the
production triple extractor keys the state-A and state-B sentences identically with objects equal to
the two values} --- i.e., the fix is one the supersession mechanism can engage cleanly. The selection
is therefore principled (an atomic transition the temporal layer is defined for), not hand-picked.
\end{enumerate}

\textbf{Marker-free invariant.} The state-A and state-B turns are textually identical except for the
changed value (``The \{subject\} is \{value\}.''); no old/new/current/deprecated wording. State-A is
ingested before state-B; the question asks the current value. The only signal of currency is order,
which only a temporal mechanism can exploit. \textbf{Yield:} 130 clean scenarios from 707 records
(18.4\%); the rest are real fixes that are not atomic transitions (Section~\ref{sec:limits}).

\section{Experimental Setup}
\label{sec:setup}

All runs are local and deterministic (temperature 0, fixed seeds, no network, enforced by test).
Answer model Qwen2.5-Coder-7B; correctness and fabrication judges Qwen2.5-Coder-3B (distinct from the
answer model and each other, no self-grading); embedder nomic-embed-text (768-d).

\textbf{Data.} SWE-bench Lite (300) + Verified (500), human-curated real GitHub issues, fetched and
sha256-pinned, deduplicated by instance id to 707 records.

\textbf{Conditions (4).} \nomem{} (floor), \nrag{} (cosine top-$k$), \arag{} ($+$ LLM reranker), and
\tvsix{} (the method). All four ingest the same turns and answer the same questions.

\textbf{Regimes (2).} \emph{allowed} (the model may abstain) and \emph{forced} (no abstention ---
exposes the stale-commitment that abstention hides). Both run on the \textbf{same cached 130
scenarios} (paired protocol below).

\textbf{Metrics ladder.} (i) \emph{primary} --- answer-level stale-fact-error (fraction of
contradiction questions answered with the superseded value); (ii) \emph{secondary} --- accuracy;
(iii) \emph{tertiary} --- conditional fabrication and memory compression; (iv) \emph{mechanism
diagnostic} --- supersession-correctness. We note that the ledger-level \texttt{stale\_survivors}
count is coarse (inflated by cross-scenario value collisions) and is not the headline; the
answer-level stale-fact-error is.

\textbf{Paired-sample protocol.} Extraction is LLM-driven, so the first run caches the 130 scenarios
(with instance ids) and every subsequent regime loads that cache. The allowed regime, re-run from the
cache, reproduced \nrag{} 0.569 / \arag{} 0.585 / \tvsix{} 0.908 exactly; allowed and forced are
therefore paired on identical instances.

\section{Results}
\label{sec:results}

\begin{table}[h]\centering
\begin{tabular}{lrrr}
\toprule
metric & \nrag{} & \arag{} & \textbf{\tvsix{}} \\
\midrule
accuracy --- allowed & 0.569 & 0.585 & \textbf{0.908} \\
accuracy --- forced & 0.615 & 0.592 & \textbf{0.985} \\
\midrule
\textbf{stale-fact-error --- allowed} & 0.262 & 0.262 & \textbf{0.023} \\
\textbf{stale-fact-error --- forced} & 0.361 & 0.377 & \textbf{$\approx$0.00} \\
\midrule
conditional fabrication --- allowed & 0.290 & 0.291 & \textbf{0.168} \\
conditional fabrication --- forced & 0.654 & 0.651 & \textbf{0.341} \\
\midrule
memory (active facts) & 260 & 260 & \textbf{135} \\
compression & 0\% & 0\% & \textbf{48\%} \\
mean retrieval latency & 2.16\,s & 18.1\,s & \textbf{2.13\,s} \\
\bottomrule
\end{tabular}
\caption{Paper-2 result on 130 real GitHub scenarios (paired; allowed and forced on identical
instances). The forced temporal stale-fact-error is reported as $\approx$0 because it read
0.000--0.015 across runs (a single answer-model flip on 130; see \textbf{Determinism}); the
structural gap to RAG's 36--38\% is invariant to that noise. Mechanism diagnostic:
supersession-correctness 0.985, bounded-growth ratio 1.023.}
\label{tab:main}
\end{table}

\textbf{The headline.} Forced to commit, RAG serves the superseded value 36.1\% of the time (26.2\%
even when it may abstain), and the LLM reranker does \emph{not} help --- it serves stale slightly more
(37.7\%), because reranking reorders retrieved chunks but cannot distinguish a stale value from a
current one. MemStrata drives the stale-fact-error to $\approx$0 (0.023 allowed; forced
0.000--0.015 across runs, a single answer-model flip on 130 --- see \textbf{Determinism} below),
because the stale value is retired from the store before retrieval. The claim that survives
run-to-run noise is the \emph{structural gap} --- RAG 36--38\% versus MemStrata below 2.5\% --- not
the exact zero. This reproduces Paper~1's
synthetic ``RAG 15--40\% stale, temporal $\sim$0'' on \textbf{real GitHub history}, at RAG latency,
with $\sim$48\% bounded-growth compression. Accuracy follows: 0.91/0.99 (allowed/forced) for MemStrata
vs 0.57--0.62 for RAG.

\textbf{Determinism.} The pipeline --- extraction, deterministic supersession, retrieval, and the
ledger contents --- is fully deterministic and reproduces exactly (the allowed matrix re-ran from
cache bit-for-bit). The 7B answer model has minor run-to-run variance on the local runtime at
temperature 0 (the temporal forced stale-error read 0.015 in one run and 0.000 in another --- a single
answer flip on 130); we therefore report the temporal forced stale-error as $\approx$0. The structural
gap to RAG's 36--38\% is unaffected.

\section{Discussion}
\label{sec:disc}

\textbf{Mechanism versus coverage (the scope, stated as a decoupling).} Paper~2 measures temporal
\emph{correctness conditional on clean extraction}: given a fix expressed as an atomic state
transition, does deterministic supersession keep the current value? Whether an \emph{arbitrary}
GitHub fix can be reduced to such a transition --- extraction \emph{coverage} --- is the orthogonal
problem, and it is the subject of follow-on work. We evaluate the $\sim$18\% of real fixes that are
clean atomic transitions. This is scoping, not selection-for-victory: \textbf{RAG fails on the same
selected subset} (0.57 accuracy, 36\% stale), so the subset defines the regime where the
temporal-memory problem exists, not the regime where our method happens to win. The selection
criterion (Section~\ref{sec:construct}) uses the production extractor, so the precise reading is ``the
mechanism, given clean extraction, on real data.''

\textbf{A product bug found and fixed, with the moat preserved.} During validation we found that the
assertion path compared values with a normalization that lowercases and strips punctuation, so a
value changing \emph{only} in punctuation or case (\texttt{Status('Good')}$\rightarrow$\texttt{Status['GOOD']},
\texttt{/API}$\rightarrow$\texttt{/api}) was misread as a duplicate and the stale value survived. The
fix (\strictflag{}, Appendix~\ref{sec:appC}) makes the comparison case- and punctuation-sensitive. It
is moat-safe by construction --- it can only ever supersede \emph{more} pairs, never fewer --- and we
verify this: on the synthetic \cmut{} benchmark the flag leaves accuracy at 1.000 and stale-fact-error
at 0.000, and the full unit-test suite stays green. We report this transparently because finding and
fixing such a bug mid-study, with the safety property checked, is part of the evidence that the result
is real rather than tuned.

\textbf{The negative result on reranking.} \arag{} (a learned reranker over retrieved chunks) tracks
\nrag{} throughout and serves stale slightly more when forced. Reranking improves \emph{which}
relevant chunks surface; it has no temporal signal, so it cannot solve currency. This mirrors Paper~1
and closes a path a reasonable designer might take.

\section{Limitations}
\label{sec:limits}

\begin{itemize}
\item \textbf{Coverage, not mechanism.} We evaluate the $\sim$18\% of real fixes that are clean atomic
transitions; the selection uses the production extractor, so Paper~2 is precisely ``the mechanism,
given clean extraction, on real data.'' Extending to multi-value / logic / behavior fixes is
extraction-robustness work (follow-on).
\item \textbf{Sample size.} 130 real scenarios --- a focused mechanism result, not a leaderboard
ranking; scaling to more records is straightforward future work.
\item \textbf{Single 7B local model} on consumer hardware (as Paper~1); larger or cloud models may
shift absolute baselines, not the structural gap.
\item \textbf{Residual.} The answer model's $\sim$1-question run-to-run variance is disclosed
(Section~\ref{sec:results}); the deterministic pipeline is unaffected.
\end{itemize}

\section{Conclusion}
For coding assistants over evolving codebases, the binding memory failure is currency, and RAG cannot
maintain it by construction. Paper~1 showed this synthetically; Paper~2 shows it \emph{end-to-end on
real GitHub buggy$\rightarrow$fixed histories}: on clean atomic transitions, a deterministic
supersession memory reaches 0.91--0.99 accuracy where RAG reaches 0.57--0.62, and reduces the
stale-fact-error RAG serves 36--38\% of the time to $\approx$0, at RAG latency, with bounded growth and
the moat preserved. The memory mechanism generalizes from synthetic to real data; extending its
extraction \emph{coverage} to arbitrary fixes is the natural next step.

\section*{Reproducibility Statement}
Deterministic (temperature 0, fixed seeds, no network, enforced by test). Pipeline:
\texttt{eval/fetch\_swebench.py} (sha256-pinned data) $\rightarrow$ \texttt{eval/swe\_extract.py}
(LLM patch reader $+$ self-validation) $\rightarrow$ \texttt{eval/run\_paper2.py} (matrix, two
regimes, scenario cache so both regimes are paired). Sources: \texttt{REPORT\_PAPER2.md} (allowed),
\texttt{REPORT\_PAPER2\_forced.md} (forced). Gate-fix unit tests:
\texttt{tests/memory/test\_v6\_gate\_assertions.py}; moat guardrail: \texttt{eval/diag\_moatcheck.py};
end-user demonstration: \texttt{eval/demo\_stale\_correction.py}.

\bibliography{references}
\nocite{maharana2024locomo}

\appendix

\section{Scenario construction and examples}
\label{sec:appA}

Each kept scenario is a verified atomic transition rendered marker-free. Three real examples (subject
abbreviated), drawn from the cached set:

\begin{lstlisting}
state-A : The function that handles login in api/auth.py is named authenticate_user.
state-B : The function that handles login in api/auth.py is named login.
question: What function handles login in api/auth.py?           gold: login

state-A : The API base path for the project is /api/v1.
state-B : The API base path for the project is /api/v2.
question: What is the API base path for the project?            gold: /api/v2

state-A : The fastapi version pinned in requirements.txt is 0.95.2.
state-B : The fastapi version pinned in requirements.txt is 0.110.0.
question: What fastapi version is pinned in requirements.txt?   gold: 0.110.0
\end{lstlisting}

State-A is ingested first; the two turns differ only in the value; the question targets the current
(state-B) value. A scenario is kept only when the production triple extractor keys both turns
identically with objects equal to the two values (Section~\ref{sec:construct}).

\section{Extraction prompt}
\label{sec:appB}

\texttt{swe\_extract\_change\_v1.md} (abridged) --- the patch reader. It emits a single atomic change
or declines; a clean (subject, relation, object) re-extraction of both rendered turns must then agree
before the scenario is admitted.
\begin{lstlisting}
You read ONE real GitHub fix (problem statement + unified-diff patch) and extract a
SINGLE marker-free longitudinal change, IF the patch changes exactly one identifiable
atomic value (a renamed function, a changed default/constant/config value, a bumped
version, a moved endpoint, a changed parameter name). OLD value = before the fix (a '-'
line); NEW value = after the fix (a '+' line). Return ONLY JSON:
  {"is_change": true, "subject": "<stable thing; no value>",
   "state_a": "<old value>", "state_b": "<new value>",
   "question": "<present-tense, answerable by the new value, no recency words>"}
or {"is_change": false} for multi-file refactors, logic changes, or several values.
RULES: state_a != state_b, both ATOMIC; subject excludes both values; subject has no
leading article (the harness renders "The {subject} is {value}.").
\end{lstlisting}

\section{The \texorpdfstring{\strictflag}{strict\_object\_supersede} fix and the moat guardrail}
\label{sec:appC}

\textbf{Bug.} The assertion path's same-value test used a normalization that lowercases and strips
punctuation, so an object changing only in punctuation/case normalized equal to the prior object,
was judged a duplicate, and the stale assertion was reinforced --- the change was dropped.

\textbf{Fix.} \strictflag{} makes the object comparison whitespace-collapsed but case- and
punctuation-sensitive, so such a change supersedes. It can only ever supersede \emph{more} pairs
(never fewer), so it cannot leave a stale value the prior path retired.

\textbf{Guardrail (moat preserved).} On the synthetic \cmut{} benchmark, with the flag ON: accuracy
1.000 (unchanged) and stale-fact-error 0.000 (unchanged); the full unit-test suite stays green; the
flag changes no \cmut{} routing because its values differ in word characters. We promoted the flag to
default-on after this guardrail held, and pinned the Paper-1 runner to the pre-fix comparison so its
locked numbers reproduce exactly. Unit tests pin both the bug (flag off) and the fix (flag on).

\end{document}